\documentclass{aastex}
\usepackage{graphicx}
\usepackage{epstopdf}
\usepackage{epsfig}
\usepackage{fullpage}
\usepackage{ulem}

\shorttitle{MultiView Methods for High Precision Astrometric Space VLBI}
\shortauthors{Dodson et\,al.}

\begin{document}

\title{
  The application of MultiView Methods for High Precision Astrometric Space VLBI at Low Frequencies \\
}
\author{R. Dodson\altaffilmark{1,7}, M. Rioja\altaffilmark{1,2},
  Y. Asaki\altaffilmark{3,4}, H. Imai\altaffilmark{5,1},
  X.-Y. Hong\altaffilmark{6}, Z. Shen\altaffilmark{6}
}
\affil{$^1$International Centre for Radio Astronomy Research, M468,
The University of Western Australia, 35 Stirling Hwy, Crawley, Western Australia, 6009;\\
$^2$Observatorio Astron\'omico Nacional (OAN), Apartado 112, E-28803,  Alcala de Henares, Espa\~na; \\
$^3$Institute of Space and Astronautical Science, 3-1-1 Yoshinodai, Chuou, Sagamihara, Kanagawa 252-5210, Japan \\
$^4$Department of Space and Astronautical Science, School of Physical Sciences, The Graduate University of Advanced Studies (SOKENDAI), 3-1-1 Yoshinodai, Chuou, Sagamihara, Kanagawa 252-5210, Japan\\
$^5$Department of Physics and Astronomy, Graduate School of Science and Engineering,
Kagoshima University, 1-21-35 Korimoto, Kagoshima 890-0065\\
$^6$Shanghai Astronomical Observatory, CAS, 200030 Shanghai, China\\
$^7$Korea Astronomy and Space Science Institute, Daejeon, Republic of Korea 305-348}

\email{richard.dodson@icrar.org}

\begin{abstract}

High precision astrometric Space Very Long Baseline Interferometry (S-VLBI)
at the low
end of the conventional frequency range, i.e. 20\,cm, is a requirement
for a number of high priority science goals. These are headlined by
obtaining trigonometric parallax distances to pulsars in
Pulsar--Black Hole pairs and OH masers anywhere in the Milky Way Galaxy and the Magellanic Clouds.
We propose a solution for the most difficult technical problems in
S-VLBI by the MultiView approach where multiple sources, separated by
several degrees on the sky, are observed simultaneously. 
We simulated a number of challenging S-VLBI configurations, with orbit
errors up to 8\,m in size and with ionospheric atmospheres consistant
with poor conditions. In these simulations
we performed MultiView analysis to achieve the required science goals.
This approach removes the need for beam switching requiring a Control
Moment Gyro, and the space and ground infrastructure required for high
quality orbit reconstruction of a space-based radio telescope.
This will dramatically reduce the complexity of S-VLBI missions which
implement the phase-referencing technique.
\end{abstract}

\section{Introduction}

Astrometry has had limited application in S-VLBI
\citep{rioja_astro_vsop,jc-pr-vsop}, as the existing missions (HALCA: \citealt{vsop}
and RadioAstron: \citealt{ra}) were designed without the rapid
beam-switching capability traditionally required for phase referencing. 
The VSOP-2 (ASTRO-G) mission was designed with this capability, but
has been cancelled due to insufficient technical readiness, mainly in
the surface accuracies of the main reflector that was required for the
highest frequencies.
The ASTRO-G design, however,  included the 
infrastructure for phase referencing, that is: the capability for
rapid source switching with attitude control (requiring a massive gyroscope) and accurate
orbit determination (requiring the application of multiple complex space-craft navigation methods).
In this paper we investigate the possibilities of using MultiView VLBI as an
alternative method to address the requirements and find that this
approach offers a solution for these aspects of phase
referencing S-VLBI on future missions.

S-VLBI is recognized as one of the most challenging areas of VLBI, as
it -- as for all space missions -- 
requires the manufacturing of complex technologies, which must be space
certified to survive long exposure to high levels of cosmic radiation,
which must be robust enough to survive the launch, which must be
reliable enough to operate without maintenance for the full life time
of the mission -- typically 3 or more years -- and yet be light enough
to not overwhelm the extremely restricted weight budget of any
satellite launch.
In addition there are the complications arising from the sensitivity
of the space-based antenna. This relate to the unfolding of a large parabolic
reflector in space whilst ensuring that the surface accuracy is
sufficiently high for observation at the highest frequencies and
downlinking a large bandwidth of coherent baseband radio data to the ground tracking stations. 
Furthermore for conventional Phase Referencing (PR) S-VLBI the
instantaneous satellite position needs to be known to a high degree
of accuracy. In \cite{a07} a careful study of the requirements for PR
observing at the ASTRO-G frequencies of 8, 22 and 43GHz concluded that
orbital reconstruction needed accuracies of a few cm. This in turn
required significant additional infrastructure both on-board and on
the ground. That is a combination of multiple spacecraft navigation
techniques consisting of an onboard GPS receiver and a Satellite Laser
Ranging Array \citep{asaki_orbit}. Whilst the engineering problem of
sensitivity is very difficult to address, new VLBI techniques can
reduce the restrictions which arise from the weight and the orbit limitations.

The rationale for S-VLBI is that it is the only method which can provide the long 
baselines that are required for the highest resolutions, where high frequencies
are not an option. Such cases would be particularly important for
astrophysical emission lines at fixed frequencies (i.e. masers), or sources for which
higher frequencies are unfeasible due to their steep spectral index
or changes in the structures under study. 

One of the best reasons for focusing on the frequencies around 1.4GHz
is that the SKA Phase-1 site decision has recently been made, with both
Western Australia and South Africa being selected to host some of the
SKA-mid (i.e. frequencies around 1.4\,GHz) dishes. The Australian site
will host a smaller array with Phased Array Feeds (PAF) to perform
un-biased surveys, whereas the SA site will host the majority of the
antennas for high sensitivity targeted observations. Both sites are
expected to be capable of supporting phased-array VLBI observations,
as this would be an extension of the pathfinder capabilities
(i.e. those of ASKAP and MeerKAT). With a phased-SKA core in Australia
and in South Africa, with the former capable of forming multiple beams
and the latter capable of forming multiple sub-arrays, high
sensitivity VLBI baselines can be formed. High sensitivity improves
astrometry, but on its own is insufficient to ensure it. With this as
our driver we are investigating how to combined the Phase 1 SKA-mid with
S-VLBI to achieve high precision astrometry.
Alternatively \cite{ed_uas_ska} reports on approaches for achieving
micro-arcsecond ($\mu$-as) astrometric accuracy with the high
frequencies that will be provided by SKA-Hi. However the time line for
the installation of SKA-Hi is not yet clear.

\subsection{L-band astrophysics with the longest VLBI baselines}

\subsubsection{Pulsar astrometry with SKA}

Pulsar astrometric VLBI is a well established technique which measures
parallaxes and proper motions, providing distances and true velocities of pulsars.
It is also a key science goal for SKA, 
as VLBI observations of any discovered Pulsar--Black Hole (PSR-BH)
candidates will be required to determine the precise distances. This
will be an essential step in the detection of gravity waves from these
PSR-BH sources \citep{strong-grav}.
These sources exist in the `Strong Gravity' regime and would allow
for the testing of predictions of quantum gravity and the search for
deviations from classical General Relativity (GR).  
The sensitivity of the SKA will lead to the discovery of many
(estimated to be hundreds) of PSR-BH pairs, of which tens are expected
to be millisecond pulsars. These, because of the requirements for
their formation, will be close to the galactic centre or in globular
clusters, that is at distances of the order of 10\,kpc. 
The full analysis for the application of all the
GR tests depends on using astrometric VLBI measurements of the
trigonometric parallax to independently determine the
distances to these sources. 
Therefore one would need to be achieving positional accuracies at the
level of 15$\mu$as at the observing frequency of 1.6\,GHz  \citep{smits_11}. This is
similar to the very best astrometric accuracies currently obtained at
wavelengths of one- to, at most, a few-cm (\citealt{reid_09} and
the references therein). 
%

\subsubsection{OH masers in the Milky Way System}

Hydroxyl (OH) masers as well as water vapor (H$_2$O), silicon monoxide
(SiO) and methanol (CH$_3$OH) masers are excellent tools for
astrometric studies. As for pulsars, trigonometric
parallaxes have been measured for 1665- and 1667-MHz (mainline) OH
masers associated with long-period variable, asymptotic giant branch
(AGB) stars \citep{langevelde00, vlemmings_03,vlemmings_07}. 
These are associated with massive-star forming regions, and evolved stars
such as AGB and post-AGB stars. Towards
the Galactic center, a large group of OH maser sources have been
detected \citep{sjouwerman_98}. Thus the OH masers are also
good astrometric tracers of the dynamics of the Milky Way.

In particular, OH maser sources in the Large and Small Magellanic
Clouds (LMC and SMC)
are excellent targets for S-VLBI astrometry. While Galactic OH maser
spots are spatially resolved in S-VLBI \citep{slysh_01} those in the LMC,
at a distance of $\approx$50~kpc, should be unresolved. There are 10 such
OH maser sources known so far
\citep{wood_86,wood_92,brooks_97,vanloon_98,brogan_04,roberts_05}. A
new OH maser survey toward the Magellanic Clouds, as part of GASKAP (Galactic ASKAP
Spectral Line Survey, \citealt{dickey_12}), will massively increase the
number of sources. 
It is expected that the proper motion of each of the Magellanic masers, which are on the order of
100\ km/s and correspond to a proper motion of $\sim$400\ $\mu$as/yr, will be
detectable. Thanks to a larger sample size of proper motions in the
galaxies ($>$100), the rotations of the LMC and the SMC may be well
modeled and the orbital motion may be unambiguously determined in an
accuracy significantly better than the 0.1 mas/yr achieved in previous measurements
(e.g., \citealt{vieira_10, kallivayalil_13}).
Furthermore, the measurement of the trigonometric parallaxes of OH masers in the LMC will
directly determine its distance, and cement the first step in the
cosmological distance ladder. To reliably detect the parallax
($\sim$20~$\mu$as, a peak-to-peak modulation of $\sim$40~$\mu$as) in
the individual maser spot motions will require astrometric accuracies 
better than this level. 

\subsubsection{Requirements}\label{sec:req}

For both science cases the aim is to achieve approximately
15$\mu$as astrometric accuracy.
Theoretically the astrometric accuracy achievable, from statistical
considerations, is approximately the synthesized beam-width over the
Signal to Noise Ratio, but in practice this is typically limited to at
most a hundredth of the beam \citep{fomalont_99}. 
The minimum L-band beam achievable with global ground baselines is of
the order of 10 milliarcseconds (mas) which, independent of any other consideration,
makes the required precision a very challenging proposition, as one would be
required to super-resolve the beam by three orders of magnitude. Any
unstable structure at these scales or source position changes 
would contaminate the astrometric results, even if the
methodological approach could in principle deliver that accuracy. The
use of weak and barely detected sources as calibrators in phase
referencing analysis will prevent
the analysis of the sources for structural changes. This is in
comparison with the well-studied ICRF sources which have been monitored for structural
variations over decades and at sub-mas scales, which is possible as they are also monitored
at higher frequencies.
S-VLBI baselines will be an order of magnitude longer than global
baselines, and would directly resolve and separate structures with an angular scale
greater than  $\sim$mas. This reduces the challenge, but does not remove
it. Nevertheless resolving to a hundredth of the beam has been
demonstrated in conventional cm-wavelength VLBI, so one would expect
the same level to be achievable with S-VLBI and  therefore provide the
astrometric levels required.

There are several approaches to deliver improved astrometric accuracies at lower
frequencies: GPS-based ionospheric corrections \citep{c_vlba_m}, wide
bandwidth corrections \citep{brisken}, in-beam corrections \citep{c_ib}
and multiple calibrator 2D corrections \citep{ef_venus,
  rioja_02, sergio}.
Most astrometric campaigns have focused on the approach followed by
\citet{c_ib} where
the dominant residual errors in the measurements are diluted by the
proximity of the calibrator to the target.
The (relative) astrometric accuracy of Chatterjee's work is 
$\sim$0.1\,mas, which is achieved by using a weak source within the
primary beam (typically $\sim$10 arc-min from the pulsar) and making
the final calibration against those. Although the proper motion and
parallax are measured to high precision, 
the ultimate astrometric accuracy is limited by the 
unknown (and unknowable) stability of
the weak in-beam calibrator.

To achieve higher precision astrometry
with a single calibrator, it must be ten times closer to improve the
astrometry ten fold.  To achieve 10$\mu$as one would therefore need a
calibrator within an arc-minute. 
Estimates from the simulations of likely detectable sources
\citep{scubed} predict that many sources will be found in-beam, but not
so many that these would have the arcmin separations required for
$\sim$15$\mu$as level astrometry \citep{godfey_ska}.
Alternatively, to achieve the SKA goals, methods that use multiple
calibrators have been proposed that interpolate solutions
to the target line of sight.  The combination of {\it
  multiple} calibrators as described by \cite{rioja_02} and
\cite{ef_venus} with the use of {\it simultaneous} in-beam
calibrators as described by Chatterjee.  Such approaches to VLBI
calibration are known as `Cluster-Cluster' or `Multi-view'
\citep{clcl,rioja09} VLBI.

\subsection{MultiView VLBI}

MultiView VLBI has been reviewed elsewhere \citep{rioja09}, and here we only
summarise the method and benefits. 
The MultiView approach is to use multiple high quality calibrators
arranged around the target, and to reconstruct the
ionospheric phase correction required in the direction of that
target. The interpolation of the required phases, accounting for
linear variations across the field, reduces considerably the need to
have the calibrators close to the target. That is the calibrators
no-longer need to be closer than 1 arc-minute for a 10$\mu$as
accuracy, as for conventional phase referencing, only that the
phase screen is sufficiently linear that the interpolated solution is
equivalent to this requirement.
The approach is most suitable for calibration of the phase residuals
from the ionosphere, which has spatial structure that is smooth
over the typical separation angle of the sources.

\begin{figure}
\includegraphics[width=14cm]{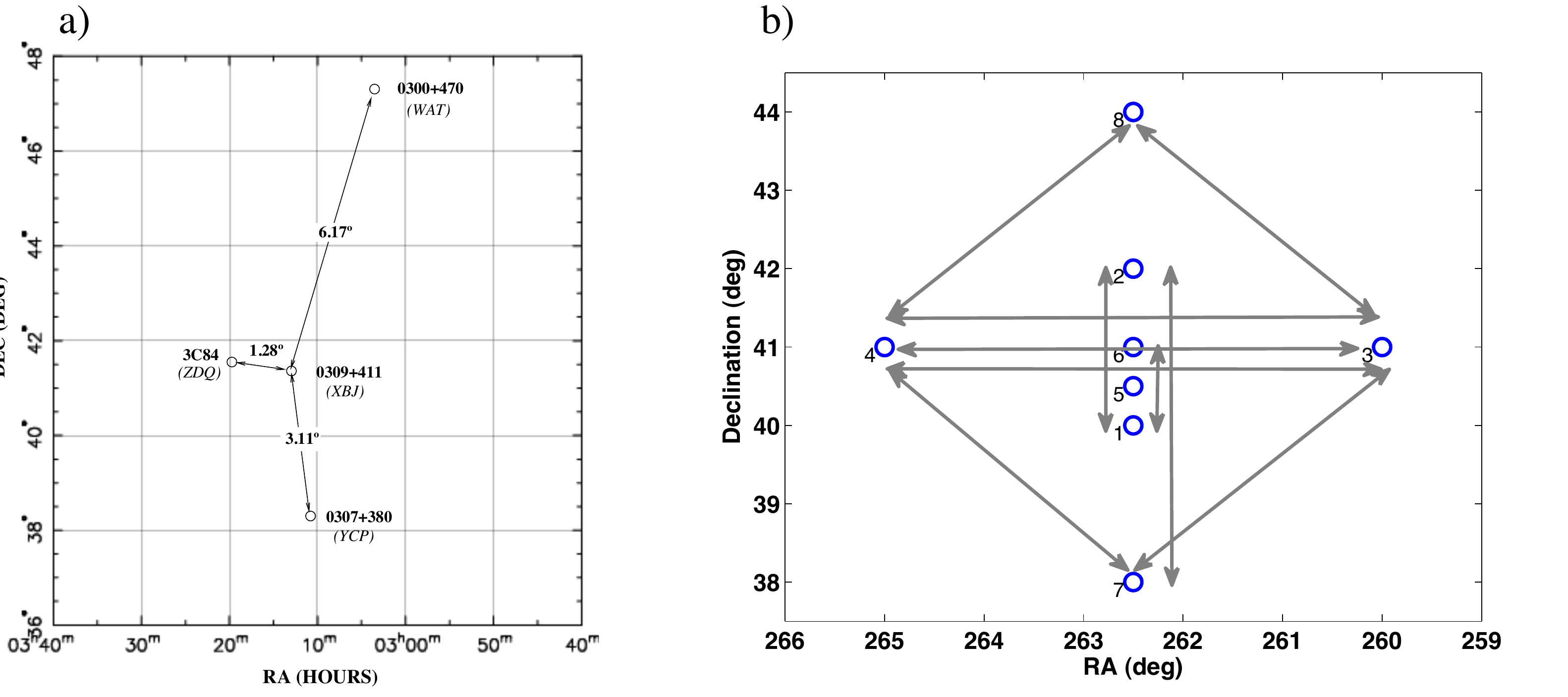}
\caption{a) Sky Coverage of the four sources targeted in the original
  Cluster-Cluster demonstration (from Rioja et al 2000). b)
  Orientation of Sources in these simulations, with the combinations
  formed indicated with light lines and triangles.}\label{fig:sky}
\end{figure}

MultiView ideally observes all the sources simultaneously so that the
temporal variations do not unduly affect the results. Nevertheless,
given that the ionospheric variations are slow, this approach works
even with fast source switching \citep{ef_venus}. We are
currently performing just such tests with the Australian VLBI Network,
the LBA, and plan to also test the method with the VLBA.
We have found, in simulations, that by using MultiView approaches we
can achieve an order of magnitude improvement in the astrometric
accuracy compared to using single calibrators, even when the MultiView
calibrators are separated by many degrees from the target.  These
simulations matched the preliminary investigations of \citet{clcl},
for which the calibrator distribution is shown in Figure
\ref{fig:sky}a. The improvement arises from the fitting of a linear
surface to the calibration residuals and interpolating this model to the
target position, which allows one to extend beyond the traditional
ionospheric patch (which is defined only by the difference between the
phase at the calibrator and that for the target). If one includes
attempts to resolve phase ambiguities one can extend the area of
validity for the solutions even further. Our approach includes this by
checking solutions which include up to a few additional wraps of phase
and determining if the new results are better than the original.

In addition this approach removes contributions from many other major
sources of calibration errors, such as the antenna clock contributions,
the antenna position error, but not source structure and position errors. 
This calibration approach is different in character to
conventional phase referencing, where the errors are diluted
by the proximity of the calibrator to the target. The limitations in
this case come from the deviation of the true phase screen from the
planar solution, which for a particular source separation will become more and more pronounced as lower
frequencies are used. We are currently attempting MultiView
calibration with long baseline LOFAR observations at 150\,MHz. LOFAR is also
capable of providing multiple tied array beams `pointed' at different
VLBI sources, and therefore will provide an interesting exploration of
the method at very low frequencies.

Previously MultiView experiments have been performed with connected
arrays where different antennas, driven by a single time standard and
under (what is approximately) a single atmosphere, are pointed
simultaneously at different targets.
As part of the VLBI Science Survey on ASKAP  \citep{tingay_vlbi} we plan to implement
MultiView methods with Phased Arrays Feeds (PAF) as installed on the
ASKAP array. 

\begin{figure}
\includegraphics[width=11cm]{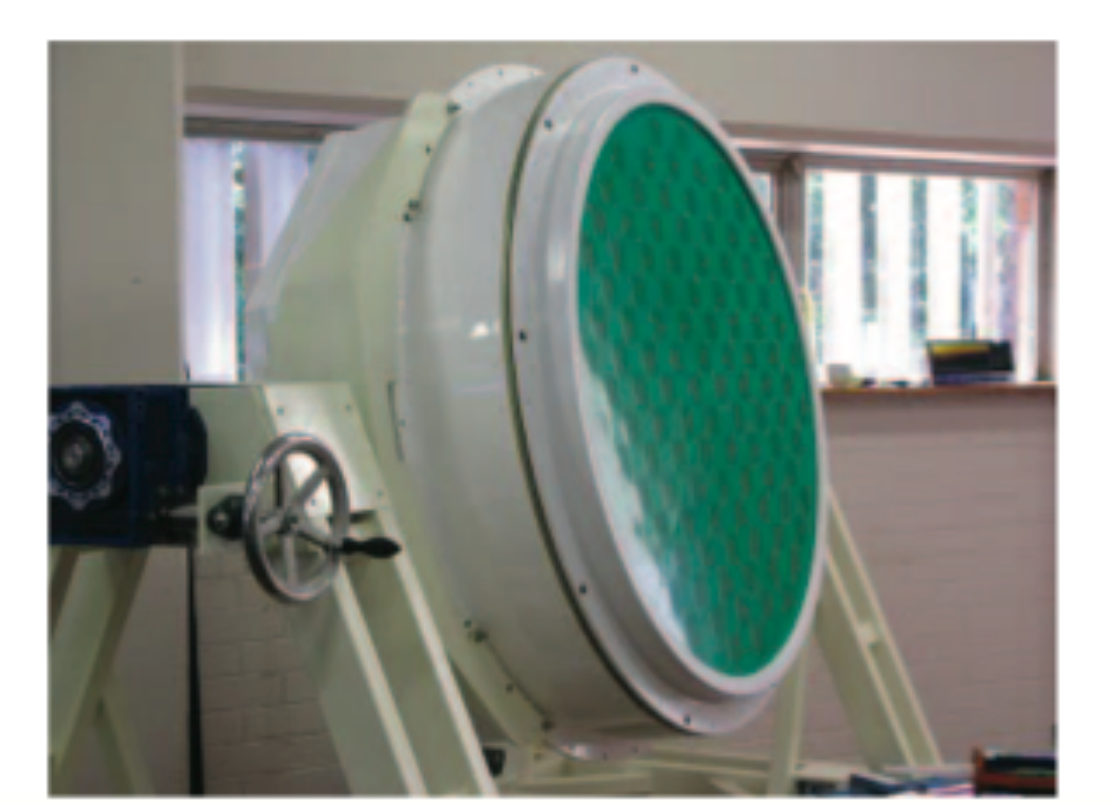}
\caption{Image of PAF from \citet{askap_paf}, the ASKAP PAF is sensitive
  between 0.8 and 1.8 GHz covering a 30 sq. deg. Field of View and  is about 2 meters across. 
}\label{fig:paf}
\end{figure}

\subsection{Phase Array Feeds in Radio Astronomy}

\subsubsection{PAF concept and PAFs on ground telescopes}

PAFs solutions for wide-field surveys \citep{paf_fisher} are being
developed for a number of instruments including ASKAP
\citep{askap_paf} and WSRT \citep{wsrt_paf}.  These devices allow
multiple beams to be formed on each antenna, extending the single
pointing of a conventional radio dish into an instrument receiving
information from multiple directions. The beams can be arbitrarily
`steered' within the Field of View (FoV) of the PAF, which is many
times the FoV of the dish, by the electronic variation of the complex
weights applied to the elements of the PAF. Figure \ref{fig:paf} shows
the PAF constructed for the ASKAP array. 

These devices are most applicable for survey projects, and the 
wide spread interest in PAF applications arise because the large field of view
optimises the `survey speed' SVS:
$ {\rm SVS} \propto N_b \Omega_b B (A_{eff}/T_{sys})^2$
where $N_b$ are the number of beams, $\Omega_b$ is the beamsize, $B$ the 
bandwidth, $A_{eff}$ the effective area and $T_{sys}$ the system
temperature. Assuming a fixed antenna size and system temperature
the best returns come about by maximising $N_b B$. 
However there is also an exciting option in the VLBI field which we are actively 
investigating; that of applying the MultiView method in VLBI with
single antennas equipped with PAFs. This forms part of the ASKAP VLBI project. In this case we are not
fully covering a blank field evenly, but pointing a smaller number of
tied beams at VLBI sources simultaneously, which are up to several
degrees apart. 

The frequency range for which PAFs will provide the best performance
is 0.5 to 15\,GHz; below 0.5\,GHz aperture arrays are more efficient,
above 15\,GHz horn arrays can produce many beams with conventional
approaches. 
ASTRON, in the Netherlands,  uses a `Vivaldi' based element array in a box pattern, forming a
travelling wave slot antenna between two coplanar conducting sheets
(e.g. \citealt{antennas}). 
The approach on ASKAP is to use a connected dipole array above a
ground plane in a `checkerboard' pattern on a printed circuit
substrate. Each patch forms two orthogonal polarisations. 
The latter approach forms an extremely compact and robust structure
and can be seen in Figure \ref{fig:paf}. We note that the PAF for
ASKAP weighs 200\,kg, but that is mainly because it is also a structural
member. The weight could be much less, given that it is based around
printed circuit boards.

\subsubsection{Phase Array Feeds in Space}

We have realised that PAFs offer interesting
possibilities in S-VLBI.  Multiview VLBI is normally considered in
terms of correcting for the approximately planar ionospheric
phase-screen across each antenna. But it also corrects for any antenna
position error in the same fashion. Offsets from the expected antenna
positions effectively add a linear phase slope across the field of
view, which is perfectly handled by the MultiView approach.

The final piece of the puzzle is the recent development of PAF
technologies, which have the potential to make Multiview VLBI a
straightforward and effective method. 
The Printed Circuit Board PAFs being developed for the ASKAP array by
CSIRO are focused on meeting the wide field of view goals of the SKA, but they are
equally suitable for MultiView VLBI. The new
generations are falling in both cost and weight and
are compact and robust.

One of the most challenging aspects of the ASTRO-G mission was to
include the capability for source switching, which is essential for
both detecting weak sources and astrometry,  using phase referencing. This
required the inclusion of a massive Control Moment Gyro (CMG).
On the other hand the electronically steerable beams which PAFs can
form would allow the tracking of multiple sources that are several
degrees apart, simultaneously,
without the CMG, providing much simpler antenna operations.

The combination of these new methods, new goals and new technologies
offers the opportunity to develop a new science mission for Space
VLBI.

\section{Simulations}

We have run simulations using the software tools ARIS
(Astronomical Radio Interferometor Simulator) and
MeqTrees. 
A detailed description of ARIS can be found in \cite{a07}. It is an
extremely complete simulator written for the ASTRO-G mission,
particularly to study the mission requirements to achieve various
science goals such as phase referencing. It allows the determination
of the required observing parameters, such as the switching cycle
time, the source separation angle, the Orbit Determination Discrepancy at Apogee (ODDA) accuracy of the satellite,
for various tropospheric conditions, and calibrator flux densities and structures.
It generates {\it uv}-data for arrays of antennas, including observational
constrains (such as low elevation limits and downlink-station
coverage). Among its many capabilities it will generate datasets which
are contaminated by the satellite orbit error, which is the capability
we have used in our simulations here. It has a simplistic ionospheric
model and, as we wished to explore more sophisticated ionospheric
options, we generated the ionospheric contamination with MeqTrees.

MeqTrees \citep{mt} is a python based interface for the CASA libraries,
and allows the construction of many different functions into
compute-trees in a straight-forward and efficient fashion. We have
used MeqTrees to generate the ionospheric models based on the 
Minimum Ionospheric Models (MIM) of \cite{mt-mim}. 
We have used the Travelling Ionospheric Disturbance (TID) and the
Kolmogorov models, as these are designed to reproduce the full range of observed
ionospheric behaviour with a limited set of parameters.

\begin{figure}
\includegraphics[width=14cm]{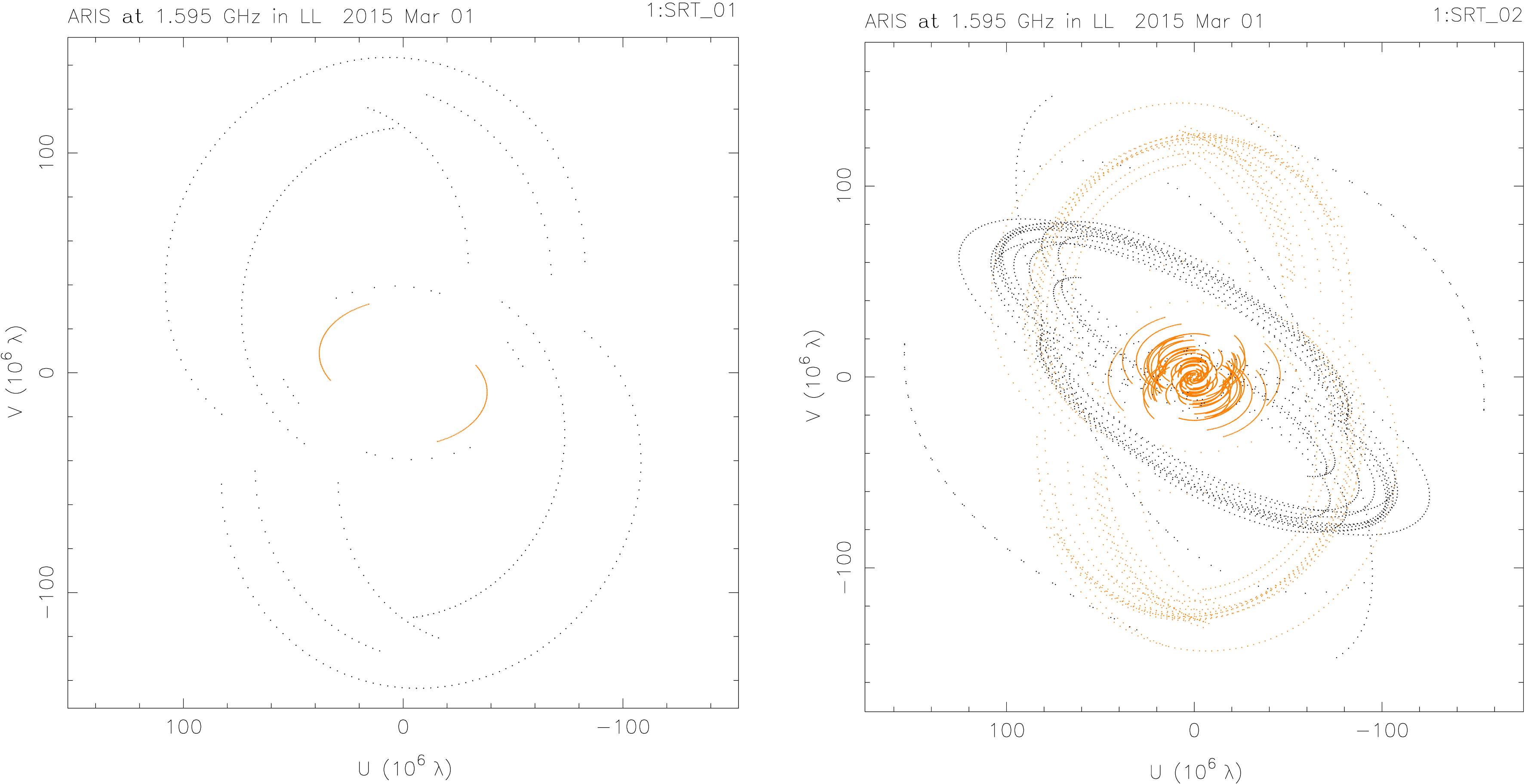}
\caption{{\it UV} Tracks for Source 1 in the simulations. All other sources
  are similar. Left) The tracks for the SKA Phase-1 model of two GRT
  and one space craft (SRT\_01), which is highlighted. 
  Right) The tracks for the full simulation with ten GRTs and two space
  craft. The second space craft (SRT\_02) are
  highlighted.
 The ground based baselines all fall
  at the centre of the plot with baselines $<$40 M$\lambda$.}\label{fig:uv}
\end{figure}

We have simulated VLBI datasets with a bandwidth of 16MHz at 1.6GHz
with two space craft in ASTRO-G like orbits, plus the VLBA antennas,
to provide a demonstration which would match an attractive
experimental setup. Twin spacecraft missions have a number of
extremely useful features, such as the absence of any atmospheric
contamination on the space-space baseline and a rapid construction of
a well-sampled uv coverage on the longest baselines. The Chinese Space
VLBI program is investigating a twin spacecraft mission, and this
drives our configuration. However the use of twin space craft has
little influence on the astrometric investigations discussed in this
paper, as they give an improved {\it uv} coverage that provides an
improved reconstruction of resolved source structure. To confirm this
we have investigated the astrometric accuracy achieved for both the
full simulation, which stands in for a model of a possible SKA-Phase 2
configuration, and a reduced dataset which closely follows the SKA-Phase 1.
Figure \ref{fig:uv} shows the typical {\it uv}-coverage in both of
these configurations.
The sources were all modelled with 1\,Jy flux and the VLBA
antennas had thermal noised added based on their nominal
sensitivities. In the discussions we estimate
sensitivities which maybe achieved in the future. 

As we are particularly interested in understanding what will be
possible in the near future with SKA Phase-1 
we selected a smaller array to realistically model the Phase-1 response. This consists of only three stations, two on the ground
and a single space craft. The antennas are separated by $\sim$9,000km
as are the two Phase-1 sites of SKA-SA and SKA-AU. 

\begin{figure}
\includegraphics[width=9cm]{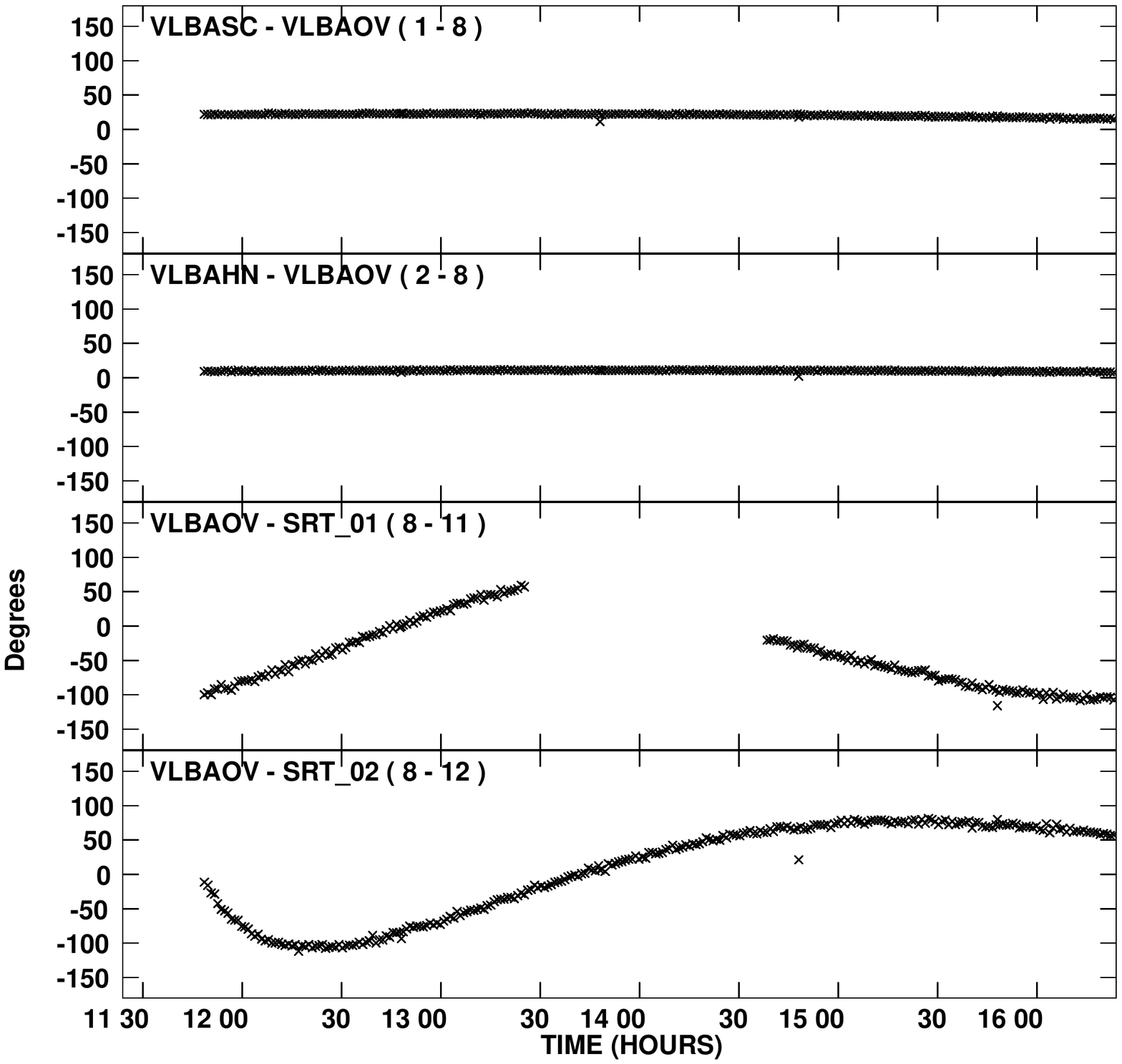}
\includegraphics[width=9cm]{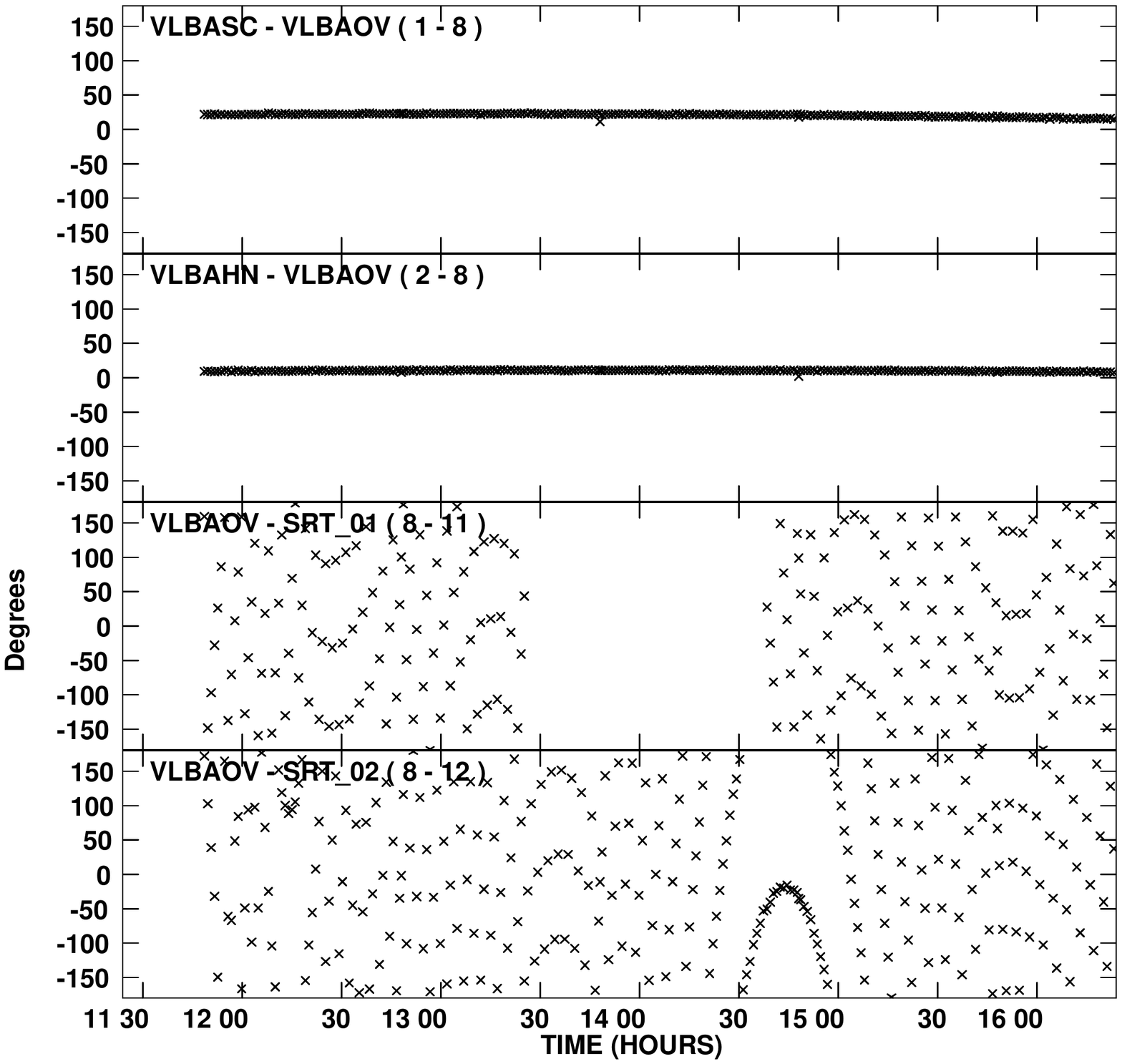}
\caption{Visibility-phase plots against time, which show  the effect
  of the positional satellite orbit errors showing 2 Ground Radio
  Telescope (GRT)
  baselines (with errors of 3mm horizontal and 10mm vertical) and the
  2 Space-Ground baselines, for a ODDA of 8cm (left) and 8m (right) 
  respectively. The data are the simulation phases for a single strong source, without
  atmospheric contamination, that is only contaminated with position
  errors.
  The data of the GRT pair baselines show a small, approximately constant,
  offset from zero whilst those of space--GRT show large
  variable deviations. For the 8m case the changes are extremely
  rapid.  }\label{fig:orb}
\end{figure}

The mission design of ASTRO-G continues to serve as a template for
S-VLBI studies. The orbit parameters of ASTRO-G were for a periastron
of 25,000 km and an orbital period of $\sim$7.5 hours. The mission design for
phase referencing required orbital errors of less than 10\,cm. This
requirement was one of the more challenging aspects for the ASTRO-G
mission.
In our simulations we have calibrated the data when contaminated by
both a ODDA of 8\,cm
and 8\,m, which are approximately the orbit error planned for ASTRO-G and
that achieved for HALCA, respectively. These two cases are shown in
Figure \ref{fig:orb}a and b respectively.

\begin{figure}
\includegraphics[width=9cm]{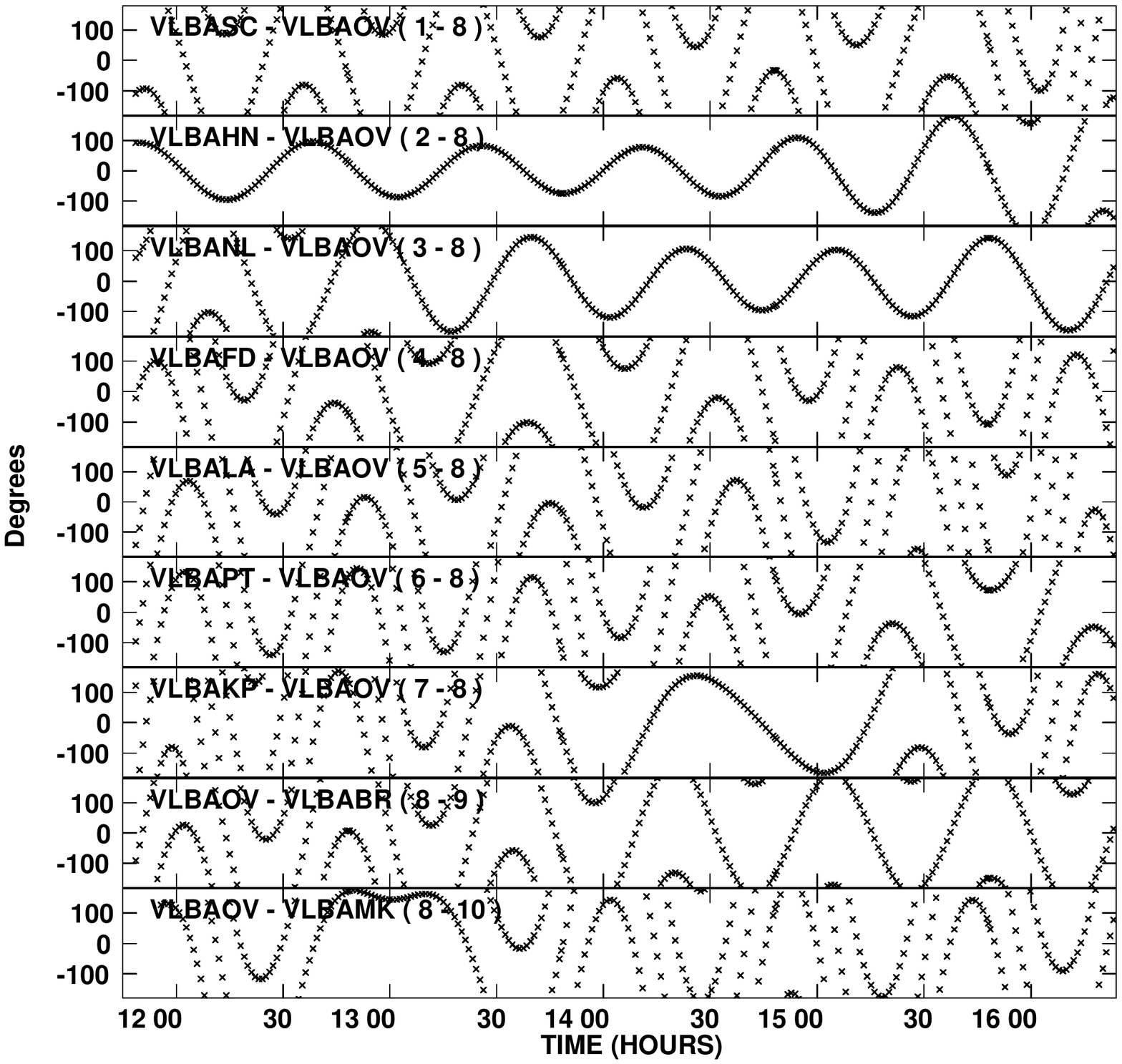}
\includegraphics[width=9cm]{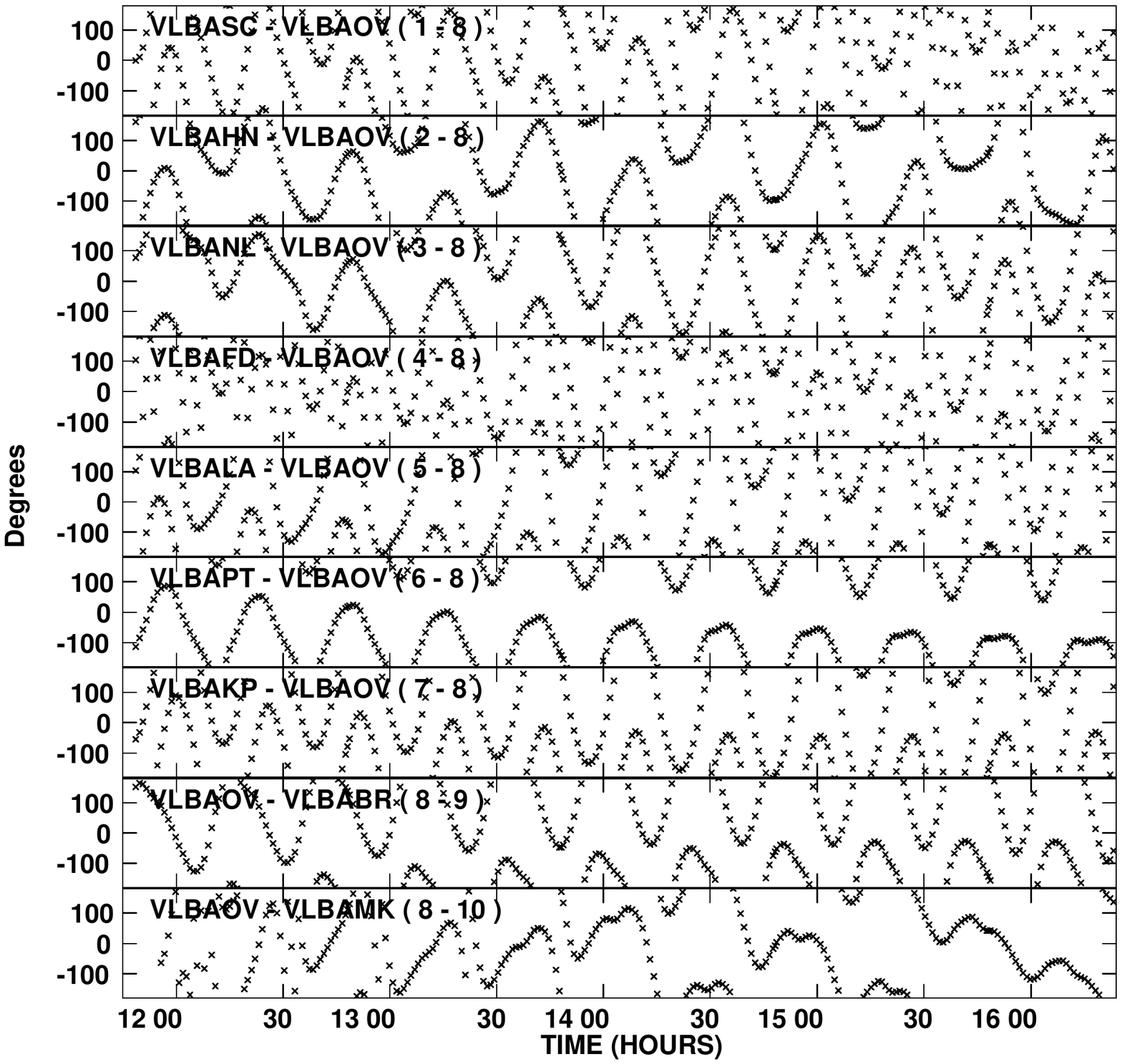}
\caption{Same as Fig. 3 but illustrating  the effect of weather
  contamination only. Shown are the simulation phases in the ground baseline
  data for a single strong source under the weather conditions
  used in the simulations: the Travelling Ionospheric Disturbance
  (left) and the Kolomogorov spectrum (right) with ASD of 6 and 12
  $\times$10$^{-13}$ respectively. See the text for the model details.}\label{fig:wth}
\end{figure}

In the simulations we used the MIM TID model with two diagonal sinusoidal disturbances, a
residual Total Electron Content (TEC) level of 5 TECU (10$^{-16}$ electrons/m$^{2}$) and a 10\% amplitude TID, at an
altitude of 200km with a velocity of 300 km/hr.  This is shown in Figure
\ref{fig:wth}a.
Our most extreme weather model was the MIM Kolomogorov power spectrum with
$\beta$=5/3, an intrinsic TEC of 10 TECU and a turbulent content of 10\%,
as shown in Figure \ref{fig:wth}b. At this level, even with a
self-calibration solution interval of one minute, the losses (13\%)
start to become significant.  Good
weather conditions are traditionally defined as having an Allan
Standard Deviation (ASD) of $10^{-13}$ at short timescales. Our models have ASD of 6 and 12$\times 10^{-13}$ at 10 seconds, respectively,
confirming that they represent extremely bad weather conditions for
the ground stations.
Other atmosphere models were explored but are not presented here, as
they do not add to the discussion.

\begin{figure}
\includegraphics[width=13cm]{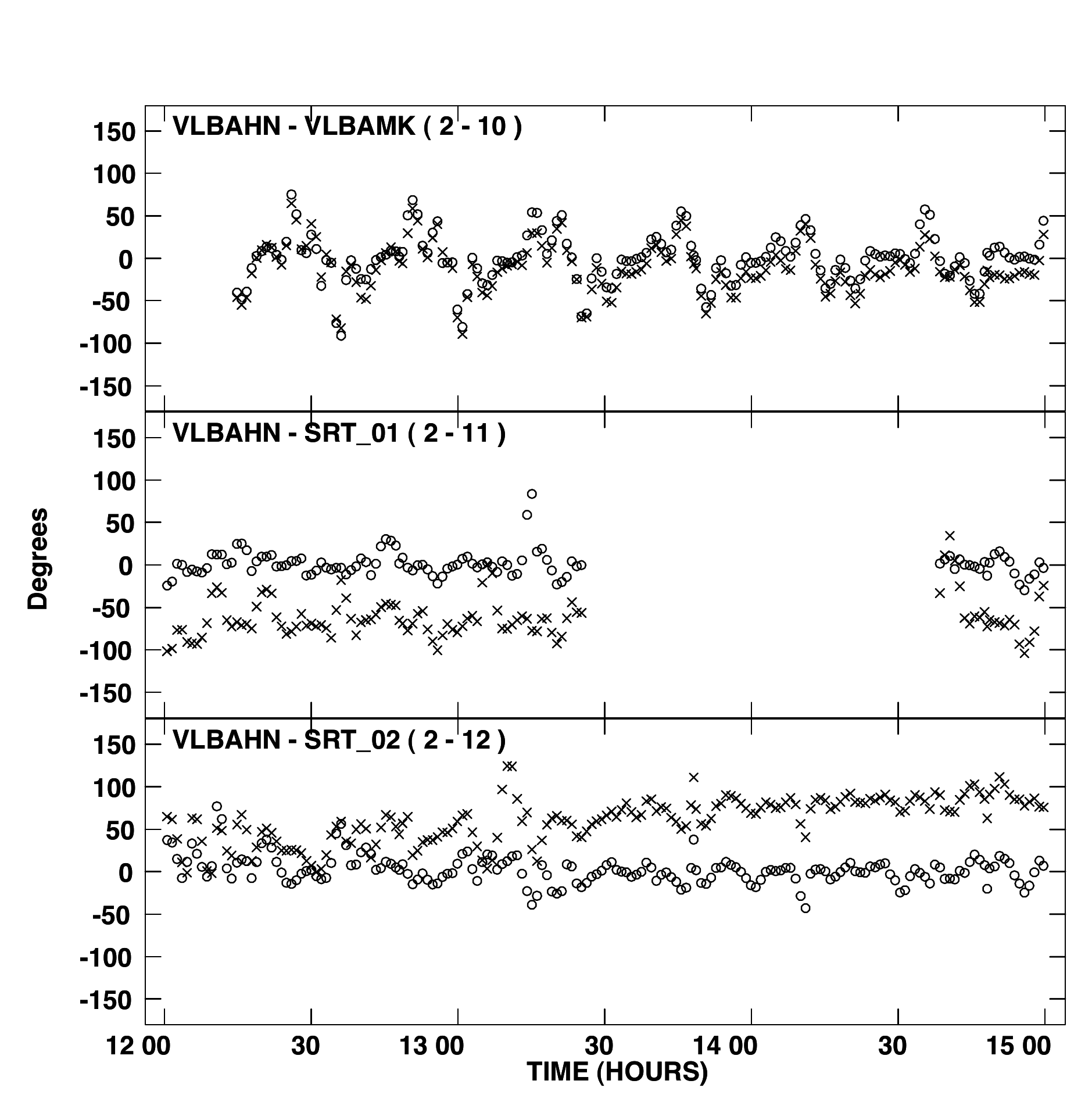}
\caption{The phase residuals for the data with large orbital
  errors and poor weather following MultiView and conventional phase
  referenced calibration. Circles mark the data
calibrated following the MultiView method and the closest calibrator
2.5$^o$ from the target (Configuration 4). Overlaid with crosses is
the same data conventionally calibrated with a calibrator having an angular separation of 0.5$^o$. This 
underlines the improvement that the 2D phase correction provides
compared to the direct phase transfer.}\label{fig:postcal}
\end{figure}

Figure \ref{fig:postcal}
shows the data with large orbital errors and poor weather,
calibrated with the MultiView method and the largest minimum angular
separation (2.5$^o$), overlaid with the same data calibrated in the
conventional manner
with the source with the smallest angular separation (0.5$^o$). This comparison
underlines the improvement that the 2D phase correction provides
compared to the direct phase transfer.

\section{Method}

We firstly simulated the {\it uv}-datasets in ARIS, with only orbital errors
and no atmosphere. 
We simulated eight sources, with the sky positions as shown in Figure
\ref{fig:sky}b. The maximum separation of the sources was 5$^o$ in RA and
6$^o$ in declination, which matches the ASKAP PAF Field of View.  Then
we converted these ID-FITS datasets to measurement sets and simulated
a common atmosphere for all sources with MeqTrees. The phases from the
calculated models 
were added to those of the orbit model. Both the TID and
Kolomogorov models were used, the latter with a range of parameters.
Each of these measurement sets were converted back into FITS format
for analysis with AIPS. 

We analysed the case of orbit only, atmosphere only and the 
combination of these in our studies. Only the analysis with all
effects are presented here. 
The residuals for each of the eight sources were measured using the AIPS task FRING and
the solution tables were exported to a text file. These were read by an external script for combination with
the correct weights for the source separation. We were particularly careful in
the phase unwrapping (as the fractional weights were not integer
values) and additionally explored the inclusion of a range of 2$\pi$
offsets around the initial results, to discover if a better solution
could be identified. In this fashion we were able to significantly
improve on the formation of the calibration via methods such as a linear combination of
interpolations using SNCOR in AIPS.
 
From these sources on the sky six configurations were formed in which
the `target' calibration was derived from the weighted combinations of
other calibrators.  These were made up of two triangular and four
linear combinations which span minimum angular separations from 0.5 to 2.5$^o$,
as listed in table \ref{tab:com}. Fourier inversion of these
calibrated datasets, including cleaning and deconvolution, was
performed with the AIPS task IMAGR.

\begin{table}
\begin{tabular}{|c|lll|cc|c|c|}
\hline
Combination \#&\multicolumn{3}{|c|}{calibrator
  IDs}&\multicolumn{2}{c|}{weight(s)}&target ID&minimum
separation (deg)\\
\hline
1&1&6 &    & 1/2 &  -     & 5 & 0.5\\ 
2&1&2 &    & 1/2 &  -     & 6 & 1.0\\  
3&2&7 &    & 3/8 &  -     & 5 & 1.5\\
4&3&4 &    & 1/2 &  -     & 6 & 2.5\\ 
5&3&4 & 8 & 1/2 & 2/3 & 2 & 2.0\\
6&3&4 & 7 & 1/2 & 5/6 & 5 & 2.5\\
\hline
\end{tabular}
\caption{Table of source combinations used, showing the calibrators involved, the
  weights in the combination, the target source and the minimum calibrator-to-target separation.}\label{tab:com}
\end{table}

\section{Results}

\begin{figure}
\includegraphics[angle=270,width=9cm]{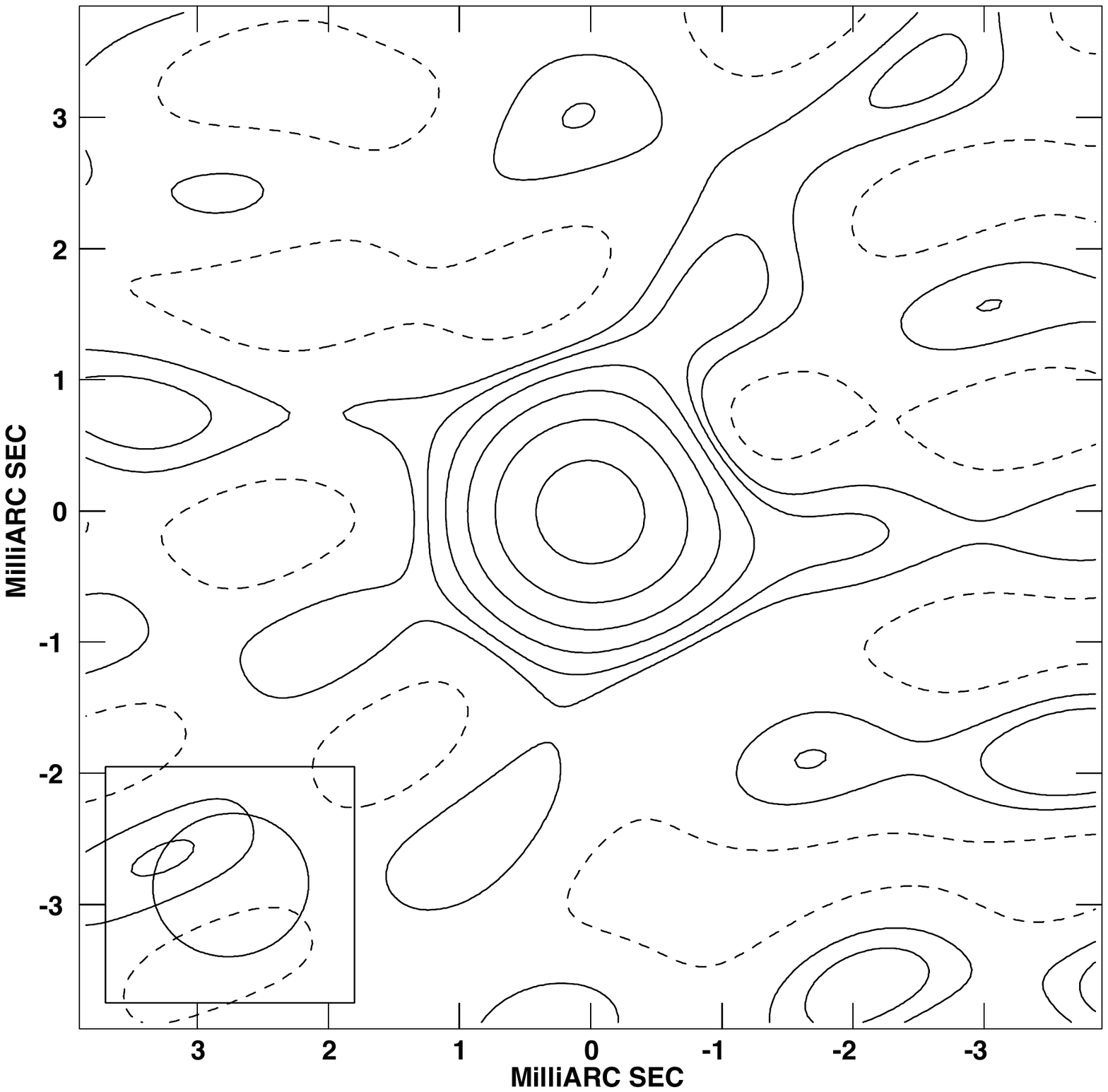}
\includegraphics[angle=270,width=9cm]{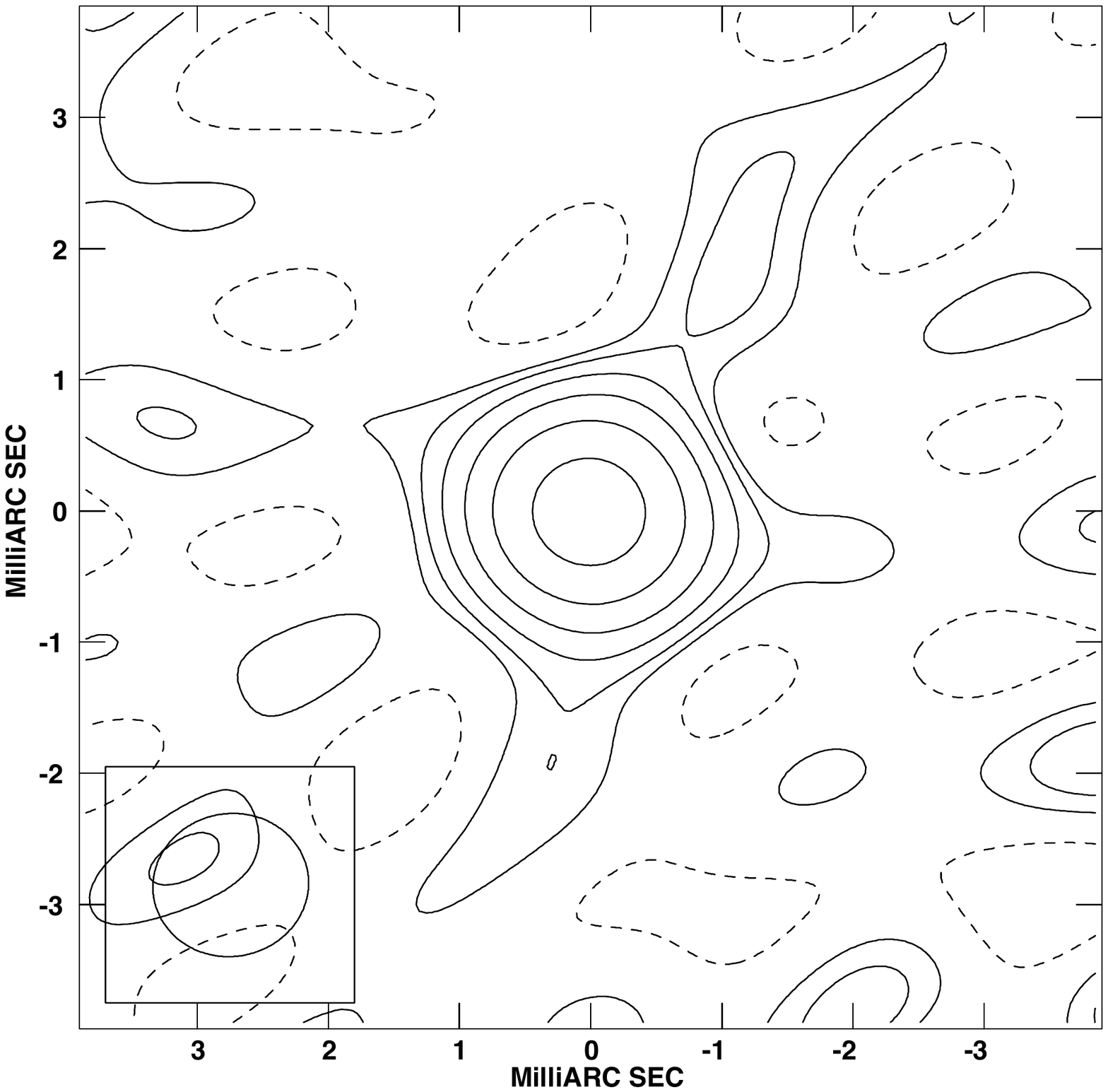}

\includegraphics[angle=270,width=9cm]{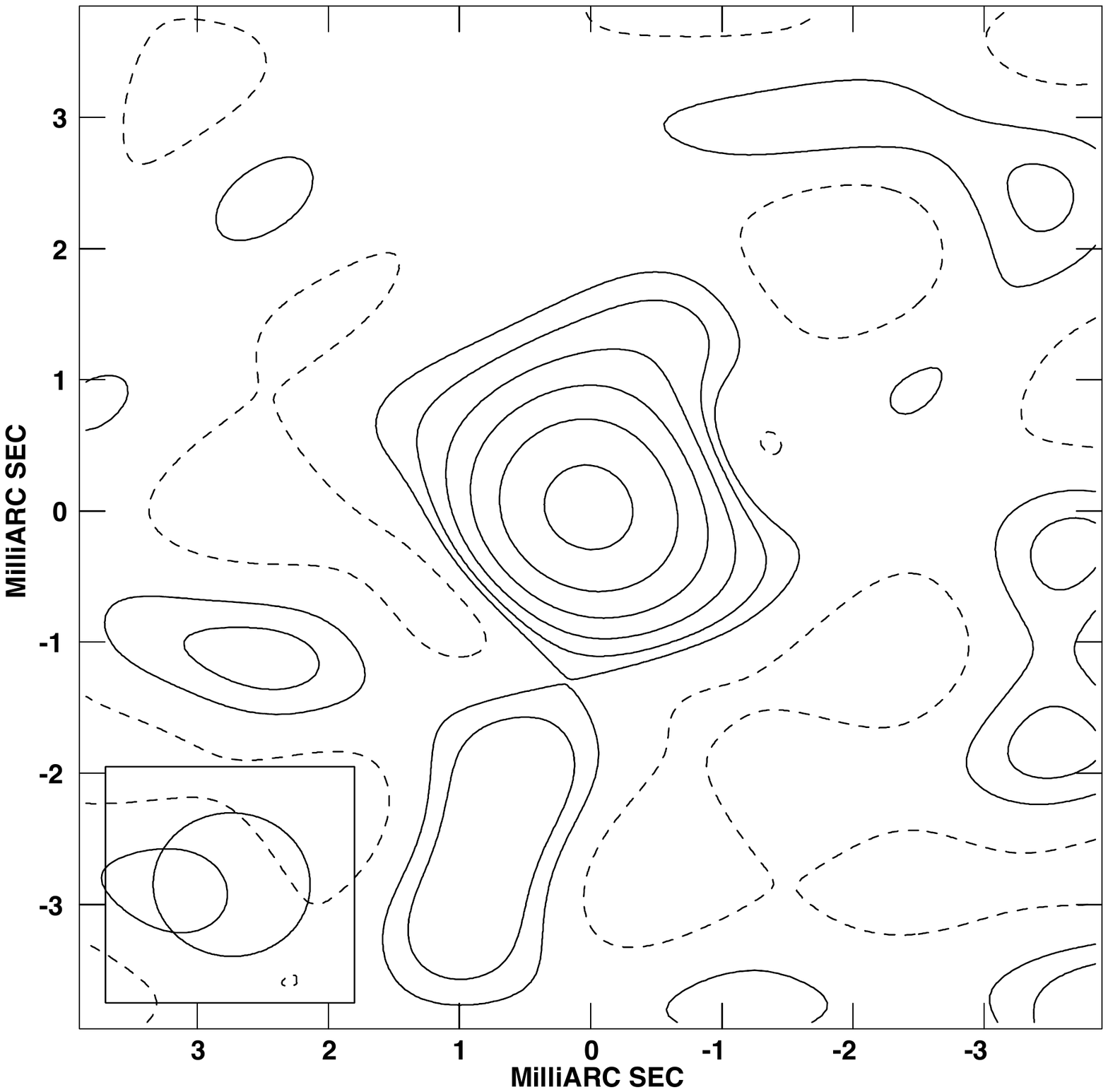}
\includegraphics[angle=270,width=9cm]{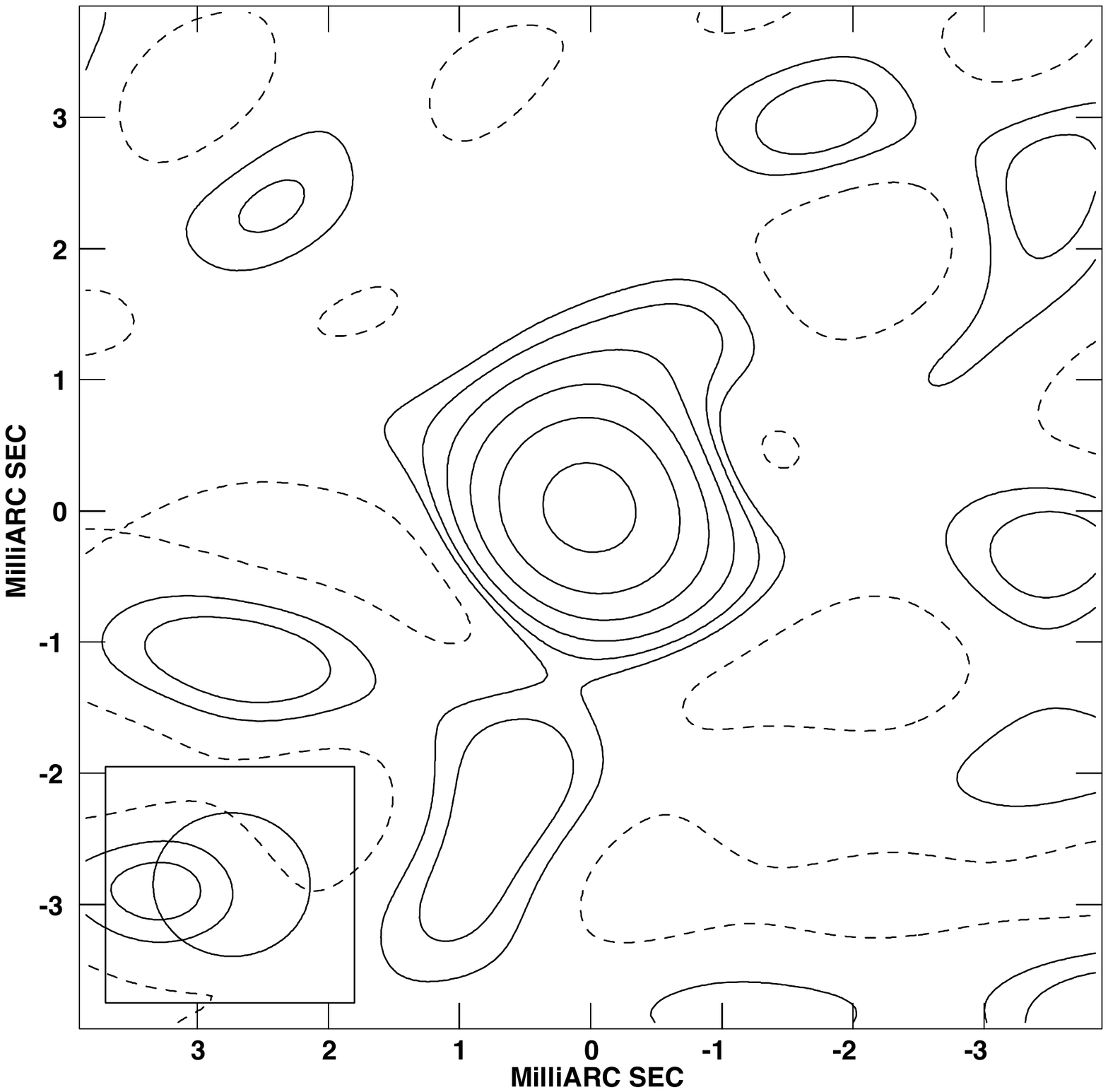}
\caption{Recovered images for the Phase-1 array using Combination 4 (Source 6 calibrated with
  Source 3 and 4)  which has the largest source separations, contaminated with the TID (top) and the Kolomogorov
  models (bottom). The simulations with a ODDA of 8cm or 8m are on the
left and right respectively. The contours run from -2,2\% and double
thereafter. The peak flux and astrometric errors are the right most
data points in Figure \ref{fig:err}}\label{fig:img}
\end{figure}

\begin{figure}
\includegraphics[width=9cm]{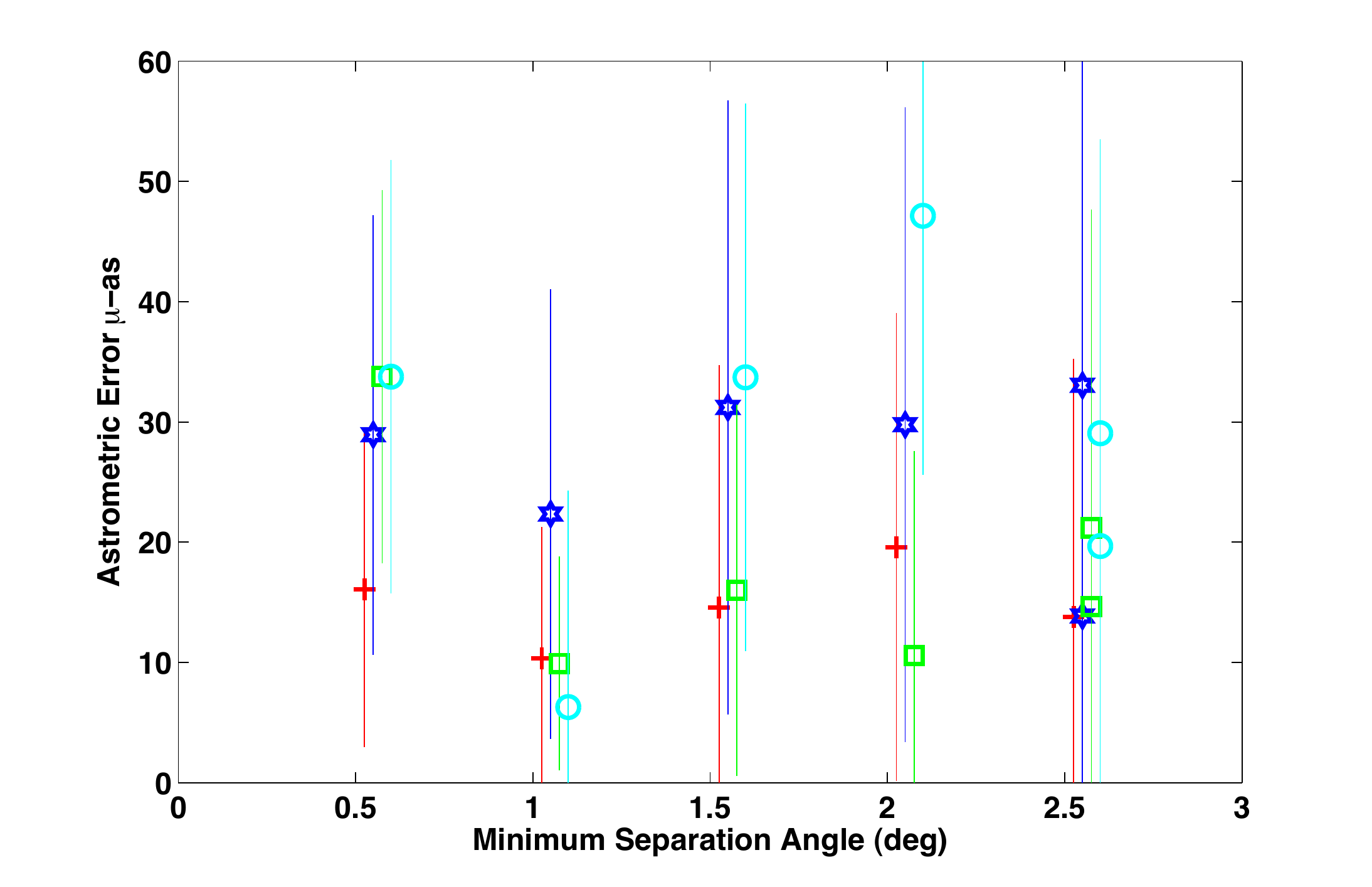}
\includegraphics[width=9cm]{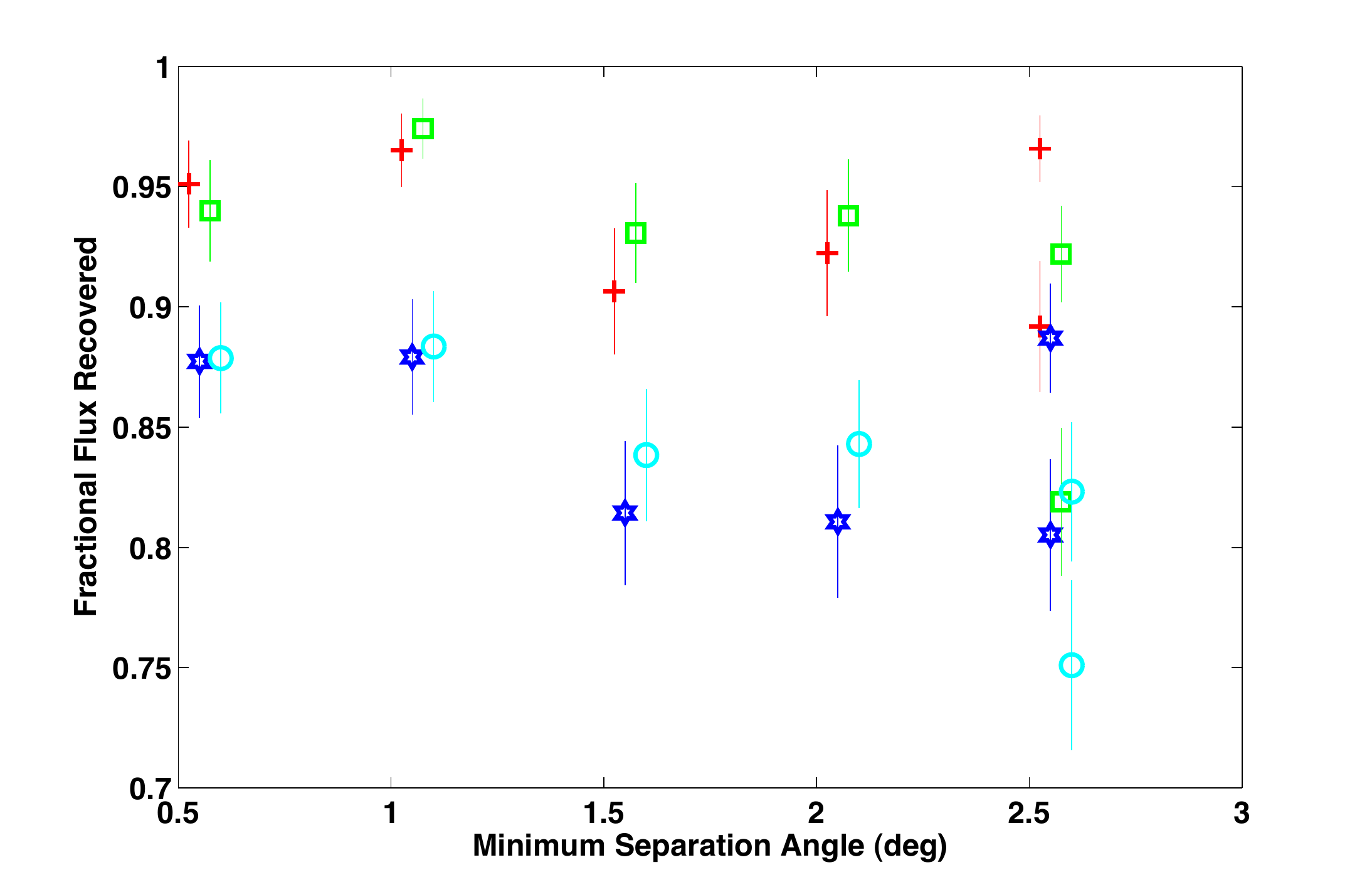}

\includegraphics[width=9cm]{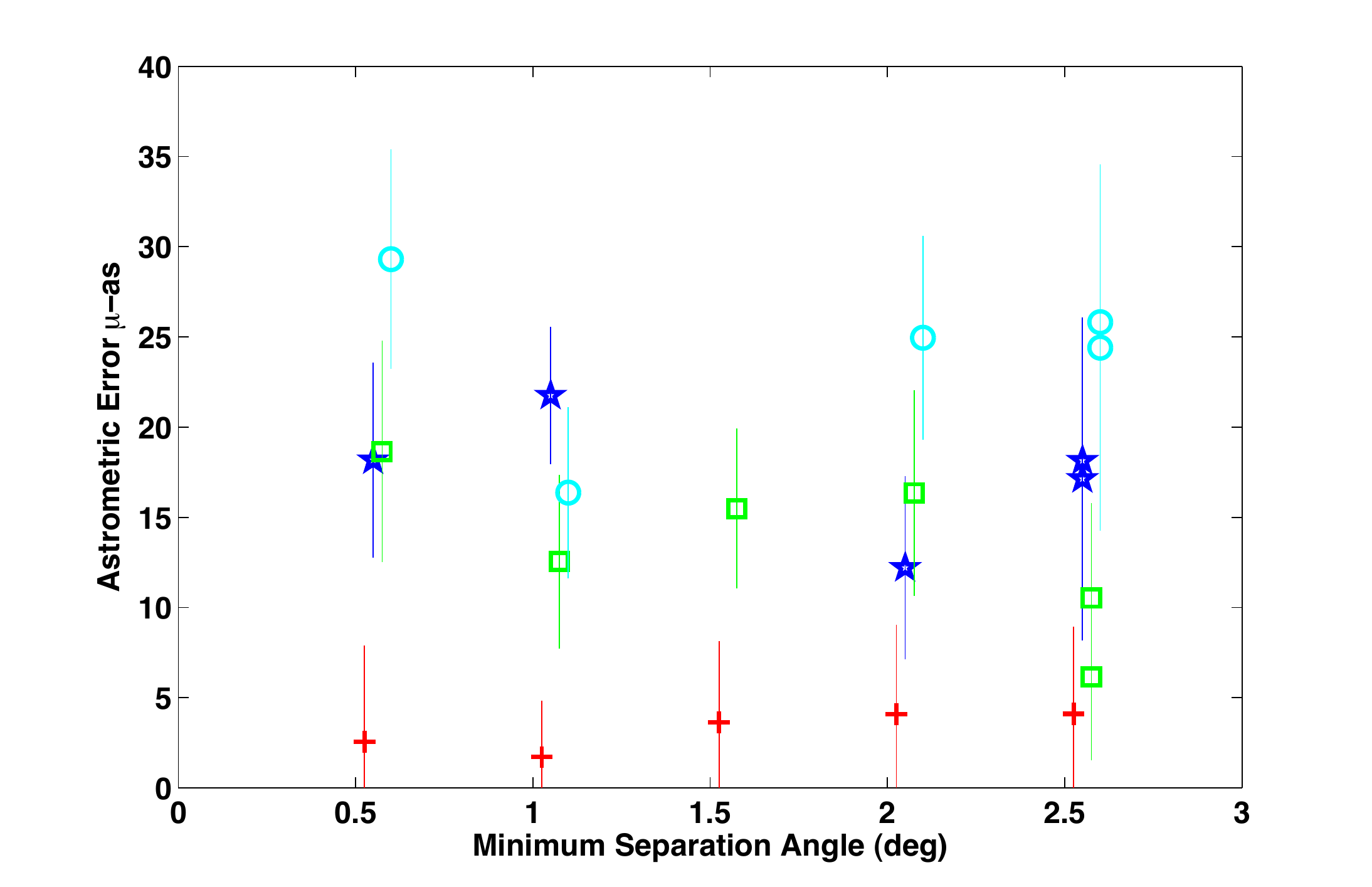}
\includegraphics[width=9cm]{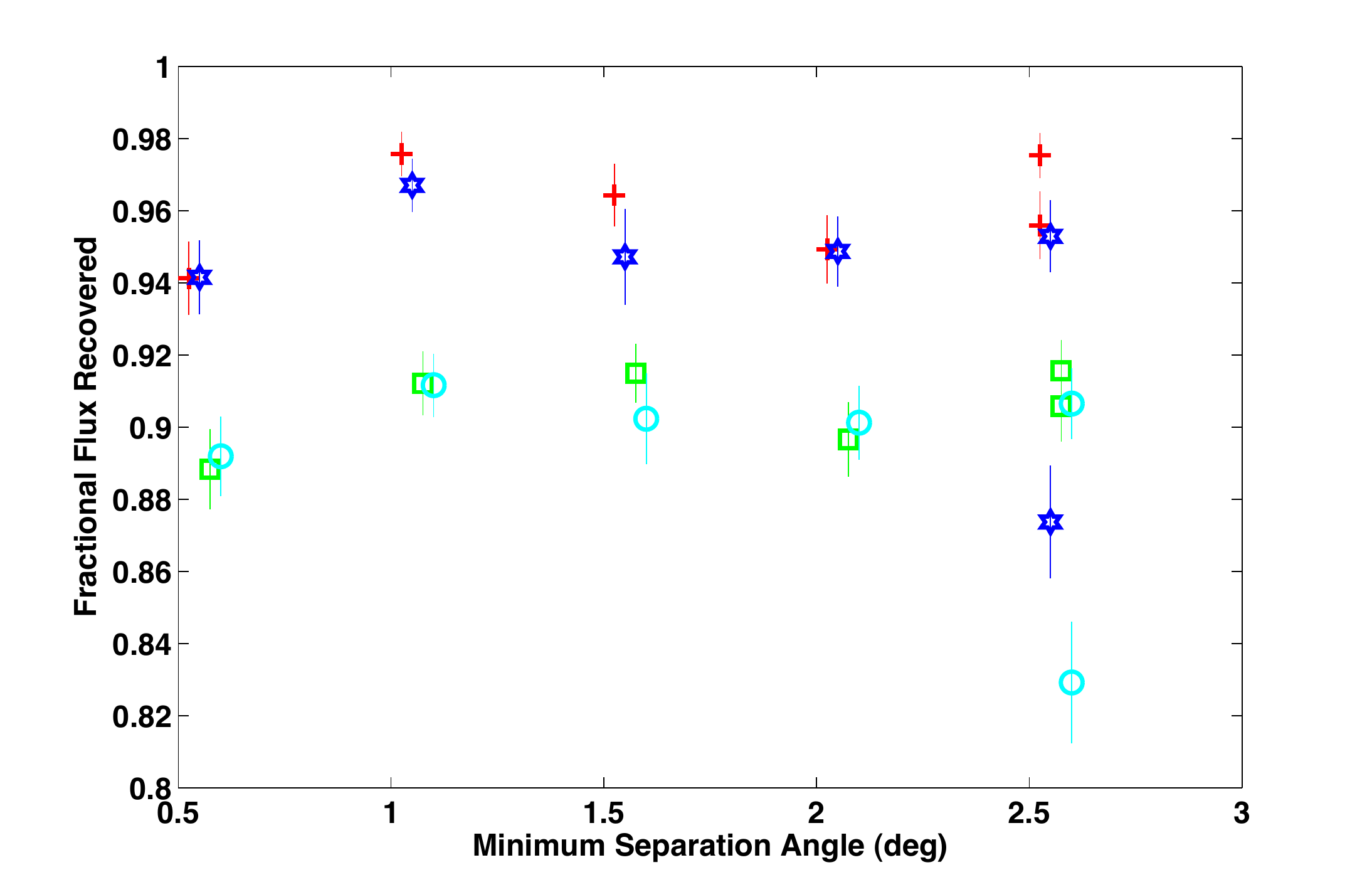}
\caption{Astrometric Errors and the fractional flux recovered in the recovered images as a function of
  the minimum calibrator distance. The upper row is for Phase-1
  simulations and the lower row for the full simulation. 
  Left) Astrometric accuracy in $\mu$-as for the four cases discussed;
  Plus signs mark ODDA errors of 8cm and the TID atmosphere, whilst
  squares mark the Kolomogorov atmosphere, stars mark 8m ODDA errors
  and TID, whilst circles mark Kolomogorov atmosphere. The error bars
  mark 1-$\sigma$ errors from the astrometric fitting. The results
  have a mean error of $22\pm10 \mu$-as for the SKA Phase-1 and
  $14\pm8 \mu$-as for the full array. Right) Flux recovered against
  minimum calibrator distance with the same symbols. The target
  accuracy is achieved except for the case of both very bad weather
  (ASD$>$10$^{-12}$) and relatively poor orbit errors (ODDA of
  8m). The different cases are slightly offset for clarity.}\label{fig:err}
\end{figure}

We have calibrated `targets' from the combinations of calibrators as
indicated in Table \ref{tab:com} and Figure \ref{fig:sky}b, and
Fourier inverted this data.  This procedure yielded images of the
target directly tied to the reference frame of the calibrators.  We
present MultiView calibrated images made with the largest calibrator
separations, for all models of ODDA and weather in Figure
\ref{fig:img}.  In Figure \ref{fig:err}a we plot the astrometric
errors against the minimum calibrator distance, for the SKA-Phase 1
configuration. In Figure \ref{fig:err}b we plot the same for the
full simulation. Figures \ref{fig:err}c and d are the same, but for the
full array. We find no correlation between the minimum source
separation and the astrometric errors. The mean astrometric error is
22$\pm$10 $\mu$as for Phase-1 and 14$\pm$8 $\mu$as for the full
array. The latter achieves our original target accuracy and the former
should be sufficient for the science goals. Table
\ref{tab:res} breaks down the errors by input ODDA and
weather and shows for the Phase-1 that the weather quality dominates,
but for the full array large orbit errors also limit the astrometry. 
This we interpret as being due to the random errors arising from
ionosphere being diluted when there are more baselines, where as
orbital errors are not reduced because they are not independent.
Similar distributions are found for the fractional recovered
flux (that is the recovered flux over model flux), with the flux
recovered and the minimum source separation being only weakly
associated. The fractional recovered flux (0.88$\pm$0.06 for Phase-1
and 0.92$\pm$0.04 for the full array) and
off-source image RMS (6$\pm$2 mJy in both cases, in line with that expected) give a
dynamic range greater than a hundred. For comparison we performed
conventional calibration for the closest pair (sources \#6 and \#5)
which have an angular separation of 0.5$^o$. For poor orbit reconstruction the
calibration fails as shown in Figure \ref{fig:postcal}, and even in
the best cases it is no better than the MultiView calibration.

\begin{table}
\begin{center}
\begin{tabular}{c|cc|cc}
&\multicolumn{2}{|c}{SKA Phase-1}&\multicolumn{2}{|c}{Full Array}\\
&\multicolumn{4}{|c}{ODDA}\\
       & 8cm & 8m & 8cm & 8m \\
\hline
TID &15$\pm$3&18$\pm$9& 3$\pm$1 &  17$\pm$3 \\ 
Kol &27$\pm$7&28$\pm$14& 13$\pm$5 & 24$\pm$5\\
\end{tabular}
\caption{Astrometric Errors (in $\mu$-as) for the four cases explored;
large and small ODDA and the two atmospheric conditions, Traveling
Wave Ionospheric Disturbances (TID) and Kolomogorov spectrum (Kol).}\label{tab:res}
\end{center}
\end{table}

\section{Discussions}

\subsection{Achieving 15 $\mu$-as astrometry with a simplified space craft}

We have demonstrated a method which allows the highest levels of
astrometric accuracy to be achieved at 1.6\,GHz. This involves the
combinations of Space VLBI to achieve small beam sizes at the
required frequency, the capability to form multiple beams
simultaneously on the antennas and MultiView analysis to solve
for the 2D phase surface. These approaches allow for the resolution of
the sub-beam effects on the mas-scale described in Section \ref{sec:req}, which would contaminate and limit the astrometric accuracies,
the removal of the massive CMG from the Spacecraft requirements
yet delivers astrometric results accurate enough to allow the measurement
of parallaxes across our galaxy and out to the LMC. 

The major advantages for the space craft from the use of MultiView
methods with PAFs are: 
i) The mission launch weight is reduced. 
ii) The source separations can be of the order of several degrees. 
iii) The orbit errors can be of the order of several meters.

We have therefore shown that the MultiView method in S-VLBI can
deliver the target astrometric accuracy of $\sim$15$\mu$-as without
the additional weight of a space borne CMG and the satellite navigation
infrastructure.
This will allow the measurement of the trigonometric parallax to PSR-BH
pairs across the galaxy and OH masers in the LMC.

\subsection{The influence of Orbit Errors}

As shown in Figure \ref{fig:err} the orbit errors contaminate the
astrometric results, raising the mean error for a 8\,cm ODDA from
$\sim$8\,$\mu$as to $\sim$20\,$\mu$as, when the ODDA is 8\,m. As shown
in Figure \ref{fig:orb}b the phase rates introduced by large orbit
errors become difficult to follow and
at this point it would become hard to track the residuals. The
HALCA orbit tracking was based on Doppler tracking at Ku-band, and
produced orbit accuracies of the order of 5\,m
\citep{rioja_astro_vsop}. A similar approach for any new mission
should therefore be acceptable.
Of course an iterative process could be used to estimate even larger ODDAs,
from the residuals, at the correlator, which would reduce the effective
ODDA to a manageable level. Such an approach could be attempted with RadioAstron.


RadioAstron was not designed for phase referencing, let alone
MultiView astrometry. Nevertheless we have considered possible
approaches which could be used.
With ground radio telescopes (GRT) alternating rapidly between the
four sources one would be able to remove their atmospheric
contamination. RadioAstron, as it has an on-board H-maser, does not
suffer from any atmospheric contamination of its own. Therefore the
cycle time between calibrators of the space craft could be
considerably longer. Even with residual orbit errors of 8\,m solution
intervals of 30 minutes allows for a 65\% flux recovery. Of course the
observing efficiency will be significantly reduced as the space
baselines will be formed only when the GRT were observing the same
source as the space antenna. 


The initial orbit errors of RadioAstron, however, are considerably greater
than that of HALCA; a recent estimate is 100\,m \citep{ed_pc}. The
errors arise from the influence of the moon on the very large orbit of
the space craft,
which prevent simple reconstruction. 
However one could perform a self-calibration to the space craft, of
the best calibrator, after GRT-only calibration. The orbit errors will
be smooth and should be easy to separate from the atmospherical
contributions. 
If these errors could be corrected for one source pre-correlation and the other
sources are within 6$^o$ (0.1 rad), the residual error will be of the
order of a tenth of this absolute error.  Therefore our results with
ODDA residuals with 8\,m should be applicable to RadioAstron.

We conclude that the orbit errors should be minimised as far as
feasible, and that traditional orbit reconstruction methods should be
sufficient for the MultiView method to provide high precision
astrometry.  Bootstrapped calibration of larger orbit errors should be
feasible, but would need to be demonstrated. 

\subsection{The effect of the SKA baselines}

We have not directly investigated the effect of the SKA baselines on the
astrometry, as that would not be significant in the setup we have
used. The sources were all modelled with 1\,Jy flux and the VLBA
antennas had thermal noised added based on their nominal sensitivities. The
SKA sensitivities are required only to detect a weak source, which we would
expect to be the target. The recommended procedure is that one uses strong ICRF
sources, which have been well monitored at a range of frequencies over
a long time, to provide confidence in the astrometric quality of the
calibrator. Both the SA and Australian SKA arrays would be capable of
supporting MultiView VLBI, either by sub-arraying or via formation of
multiple outputs from their PAFs, along with conventional connected
arrays such as VLBA, ATCA, GMRT and WSRT.

With the SKA site decision now announced we can make estimates of the
Phase-1 sensitivity. SKA-AU will have 96 12m-antennas with PAF
receivers 
whilst SKA-SA will have 254 single pixel 13.5m
antennas. 
Assuming system temperatures of 30K and 20K respectively and that the
SKA-SA antennas would be evenly sub-arrayed between the four sources
for observation we find that these station beams have similar
sensitivities of $\sim$500 m$^2$K$^{-1}$. For a bandwidth of 256MHz
and a one minute solution interval such a station correlated with 10m
diameter antenna with a system temperature of 50K would provide a
baseline sensitivity of the order of 0.3 mJy. This underlines that for
Space-VLBI, even when combined with the SKA, will struggle to detect
the postulated weak in-beam calibrators; a consequence of the small
diameter of any antenna capable of being launched into space.  Note
that this mJy limit will only be for the calibrators not the target,
which would have an integration time of the whole observation. ICRF
sources would fullfil these requirements with ease.  It is also worth
commenting that for Phase-1 VLBI large telescope, such as the
Parkes 64m, will continue to provide significant sensitive baselines.

We conclude that S-VLBI missions can be linked to the SKA ground
stations and we believe that this will be the only method that can
produce the high precision astrometry required for the fulfillment of
the SKA key science goal of the measurement of trigonometric parallax
distance to PSR-BH targets.


\end{document}